\documentclass[twocolumn,prl,showpacs,amsmath,amssymb]{revtex4}
\usepackage{amsmath}
\usepackage{dcolumn}
\usepackage[dvips]{graphicx,color}
\usepackage{wrapfig}

\newcommand{\bra}{\left\langle}
\newcommand{\ket}{\right\rangle}

\newcommand{\p}{\partial_t}
\newcommand{\e}{{\rm e}}

\newcommand{\hc}{h_{\rm c}}
\newcommand{\Dt}{\Delta t }
\newcommand{\thetas}{\theta^{\rm s}}
\makeatletter
\begin{document}

\title{Critical phenomena in globally coupled excitable elements}

\author{Hiroki Ohta}
\email{hiroki@jiro.c.u-tokyo.ac.jp}
\author{Shin-ichi Sasa}
\email{sasa@jiro.c.u-tokyo.ac.jp}
\affiliation
{Department of Pure and Applied Sciences,
University of Tokyo, 3-8-1 Komaba Meguro-ku, Tokyo 153-8902, Japan}
\date{\today}


\begin{abstract}
Critical phenomena in globally coupled excitable elements 
are studied by focusing on a saddle-node 
bifurcation at the collective level. Critical 
exponents that characterize divergent fluctuations of 
interspike intervals near the bifurcation are calculated 
theoretically. The calculated values appear to be 
in good agreement with those determined by numerical 
experiments. The relevance of our results to jamming 
transitions is also mentioned.
\end{abstract}

\maketitle

The understanding of cooperative phenomena in nonequilibrium systems
is one of the most important problems in physics. In contrast to those
in equilibrium systems, their  nature  essentially depends on the 
dynamical properties of the systems. This makes it difficult to 
perform a systematic study of the problem. Thus, it is necessary to 
investigate the typical cooperative phenomena in nonequilibrium systems.
 
Recently, critical behaviors have been observed experimentally 
\cite{Segev1,Plenz1,Segev2,Bedard,Diakonos,Yamamoto,Harada} 
and numerically \cite{Sakaguchi1,Herrmann1, Beggs,Kinouchi,Herrmann2,Teramae,
Buice,Capano} in typical examples of coupled excitable elements 
such as neural networks and cardiac tissues. In general, 
such critical behaviors are classified into several groups on the basis of the 
exponents of divergences.
The classification enables the realization of a universality class 
for critical phenomena in coupled excitable elements. However, the broad 
distribution of the phenomena makes it difficult to elucidate the mechanism 
of the criticality. When we consider the role of the mean-field Ising 
model in theories of equilibrium statistical mechanics, 
it becomes apparent that it is necessary to develop an elegant method for 
theoretical analysis of a minimal model describing critical phenomena in 
coupled excitable elements. 

Thus, with this aim in mind, 
we analyze a previously proposed simple model for 
globally coupled excitable elements in this Letter \cite{Shinomoto}. 
In particular, we focus on a divergent
 behavior with respect to parameter change 
around a {\it saddle-node bifurcation} because 
the excitability of this model is related to the bifurcation.
It should be noted that such transition properties have been studied 
for different excitable systems \cite{Sakaguchi1,Buice,Plenz1,Beggs,Kinouchi}.
The main achievement of this Letter is a theoretical derivation of 
the critical exponents that characterize the singular behavior 
near a saddle-node bifurcation.

\paragraph{Model:}                           %

\begin{figure}
\includegraphics[width=7cm,clip]{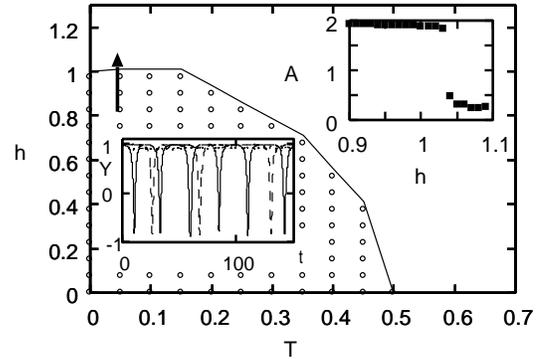}
\caption{Phase diagram. $\bra L\ket$ on the displayed curve satisfies 
$0.097< \bra L \ket <0.107$. 
Inset: (right) Amplitude of oscillation A as a function of $h$. 
(left) Typical samples of the time evolution of $Y$ for $h=1.0$ (solid line), 
$h=1.02$ (dashed line), and $h=1.1$ (dotted line). Here, $T=0.05$ and $N=100$. 
The arrow represents the direction of the parameter change.
$\bra L \ket > 0.1$ in the states where the symbols are placed.}
\label{sig}
\end{figure}

The excitable nature of a system is characterized by the existence
of spikes in a time series. Mathematically speaking, spikes are
described by trajectories near a homoclinic orbit in a differential
equation. As a simple example,  let us consider an ordinary differential 
equation $\partial_t \phi= \omega- h\sin\phi$ for a phase variable 
$\phi \in [0,2\pi]$, in which there exists the homoclinic orbit when 
$\omega=h$. Then, when  $h$ is slightly larger than $\omega$, a small 
perturbation for the fixed  point $\phi_* =\sin^{-1}({\omega/h})$ yields
one spike. On the other hand, when $h $ is slightly less than $\omega$, 
the system shows an array of spikes with a long interspike interval.
The qualitative change in the trajectories is an example of saddle-node 
bifurcation. By using this simple dynamics as a model of excitable element, 
we study globally coupled excitable elements 
$\{ \phi_i \}_{i=1}^N$ under the influence of noise \cite{Shinomoto}:
\begin{align}
\partial_t \phi_i &= \omega -h\sin\phi_i-\frac{K}{N}\sum_{j=1}^N
\sin(\phi_i-\phi_j)+\xi_i,
\label{model}
\end{align}
where $\xi_i(t)$ is Gaussian white noise that satisfies the relation
$\bra \xi_i(t) \xi_j(t') \ket=2T \delta_{i,j}\delta(t-t')$. 
Without loss of generality we can assume $K=1$, and we restrict 
our investigations to the case $\omega=1$. The control parameters 
are $h$ and $T$. All numerical results in this Letter have been 
calculated by employing an explicit discretization method with 
a time step $\delta t=0.05$.

The collective behavior of this system is described by the time 
evolution of a complex amplitude, which is given by 
\begin{equation}
Z\equiv \frac{1}{N}\sum_{j=1}^N \e^{i \phi_j}.
\label{Zdef}
\end{equation}
In particular, for the expression $Z=X+iY$, where $X$ and $Y$ are real numbers,
 the expectation values of the angular momentum 
$L\equiv X (\partial_t Y) - (\partial_t X ) Y $ are used to distinguish 
the oscillatory states ($\bra L \ket \not =0$) from the stationary states 
($\bra L \ket =0$)
\cite{Kuramoto3}. In Fig. \ref{sig}, we show an approximate phase diagram 
in the form of a curve that satisfies the condition 
$0.097 < \bra L\ket < 0.107$ in the parameter space $(T, h)$.
A similar phase diagram was obtained in Ref. \cite{Shinomoto} 
by measuring the frequency of the time-dependent 
phase distribution function. 
Here, the curve starting from $(T,h)\simeq (0, 1)$ is related to a 
saddle-node bifurcation, while the curve from $(T,h)\simeq (0.5,0)$ 
is related to a Hopf bifurcation.
It should be noted that a complicated bifurcation diagram appears near $T=0.1$, 
which originates from the Takens-Bogdanov type bifurcation \cite{Sakaguchi2}. 

\paragraph{Preliminary:}                    %

In this study, we focus on systems near a saddle-node bifurcation.
First, we fix $T=0.05$ and change $h$ across the bifurcation 
from below. Here, the amplitude of oscillation 
$A\equiv \bra [{\rm max}_t X(t)  -{\rm min}_t X(t)] \ket $ changes 
discontinuously at the bifurcation. (See inset of Fig. \ref{sig}.) 
The discontinuous change in the amplitude is in a sharp 
contrast to a super-critical Hopf bifurcation at a collective level, 
where the amplitude of oscillation changes continuously at the 
bifurcation \cite{Kuramoto2}.
It should be noted that in a manner similar to that of the critical 
phenomenon in equilibrium systems, the continuous transition leads to
a critical divergence of amplitude fluctuation \cite{Daido}. 
(See also Refs. \cite{Ritort,Strogatz2} as reviews.)  
Thus, the discontinuous nature of the transition is not
indicative of the appearance of critical phenomena.

\begin{figure}
\includegraphics[width=6cm,clip]{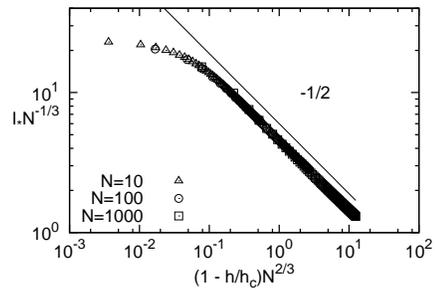}
\caption{$I_*N^{-1/3}$ as a function of $(1-h/\hc)N^{2/3}$. Here, 
$\hc=1.0283$. 
The guide line represents a power-law function with exponent $-1/2$}
\label{i}
\end{figure}

\begin{figure}
\includegraphics[width=6cm,clip]{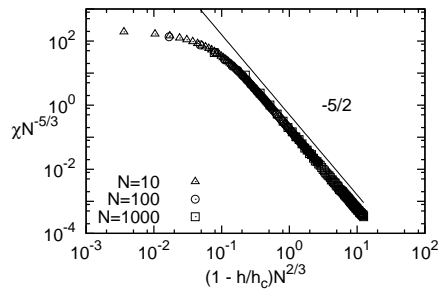}
\caption{$\chi N^{-3/2}$ as a function of $(1-h/\hc)N^{2/3}$. 
Here, $\hc=1.0283$.
The guide line represents a power-law function with exponent $-5/2$}
\label{fi}
\end{figure}

Nevertheless, based on the fact that a typical time scale diverges at a 
saddle-node bifurcation, we take into account the fluctuation of interspike intervals. Explicitly, by using the phase of collective 
oscillation,
\begin{equation}
\theta\equiv{\rm arg}(Z),
\label{thetadef}
\end{equation}
we define the interspike interval $\hat I$ 
as the minimum time interval $[t,t+\hat{I}]$ 
over which the time integration of $\partial_t \theta$ is equal to $2\pi$ 
for a time $t$ satisfying $\theta(t)=-\pi/2$.
As the most primitive statistical quantities of $\hat I$, 
we measured its average and fluctuation intensity defined by
\begin{align}
I_*(h,N) &\equiv \bra \hat I \ket,\label{bi} \\
\chi(h,N)&\equiv N \left( \bra \hat{I}^2 \ket- \bra \hat{I} \ket^2 \right).
\label{bc}
\end{align}
In order to determine the divergent behaviors near the bifurcation 
in the thermodynamic limit, we performed finite-size scaling 
analysis by using systems with $N=10$, $100$, and $1000$.
For each system, the values of 
$I_*(h,N)$ and $\chi(h,N)$ were calculated for several values of $h$. 
Then, we assume the scaling relations
\begin{eqnarray}
I_*(h,N)
&\simeq&
N^{\frac{\zeta}{\nu}}
F_{I}
\left(
\frac{\hc-h}{\hc}
N^{\frac{1}{\nu}} 
\right), 
\\ 
\chi(h,N)
&\simeq & 
N^{\frac{\gamma}{\nu}}
F_{\chi}
\left(
\frac{\hc-h}{\hc}
N^{\frac{1}{\nu}}
\right),
\end{eqnarray}
where the exponents $\nu$, $\zeta$, and $\gamma$ and the critical value $\hc$
are determined so that the scaling relations are valid. We also assume
that a distribution function of $\hat I$ is 
expressed as a function of $\hat I N^{\zeta/\nu}$
when $h=\hc$. By applying this assumption to $\chi(h,N)$ in (\ref{bc}), 
we find a relation $\gamma/\nu= 2\zeta/\nu+1$, which yields
\begin{equation}
\nu=\gamma-2 \zeta.
\label{scaling}
\end{equation}
Moreover, since $I_*$ and $\chi$ are independent of $N$ in 
the regime $(\hc-h)N^{1/\nu} \gg 1 $, the asymptotic
behaviors can be derived as $F_I(x)\simeq x^{-\zeta}$ and 
$F_{\chi}(x) \simeq  
x^{-\gamma}$. With the consideration of these conditions, 
we determine  the values  $\hc=1.0283$, $\nu= 3/2$, $\zeta=1/2$, 
and $\gamma=5/2$, for which the excellent collapses 
to universal curves are found, as displayed 
in Figs. \ref{i} and \ref{fi}.

\paragraph{Theory:}                          %

We now present a theory for the results $\zeta=1/2$ 
and $\gamma=5/2$. ($\nu$ is then determined from (\ref{scaling}).)
In the argument below, we assume that 
$\epsilon\equiv \hc-h$ is a sufficiently small positive constant 
and consider the asymptotic limit $N \to \infty$ for the assumed 
value of $\epsilon$. 

We first notice that for a sufficiently small value of $T$, 
the excitable elements are almost in synchronization. 
Thus, when setting 
$\phi_i=\theta+\delta\phi_i$, we assume that $|\delta\phi_i| \ll 1$.
From this assumption and the definition of $\theta$ given in 
(\ref{thetadef}), we can derive the equation 
\begin{align}
\partial_t\theta = \omega - h \sin \theta + \eta,
\label{evol:theta}
\end{align}
with  $\bra \eta(t)\eta(t')\ket = 2 (T/N) \delta(t-t')$,
where we have ignored the contribution of 
$O( \sum_{i=1}^N (\delta \phi_i)^2/N)$
to the time evolution of $\theta$. Within  this approximation,
$\hc$ is determined as $\hc=\omega$. 
Although the equation we analyze has become quite simple, 
the calculation of the critical exponents is still non-trivial.
By using a special technique, we can derive the distribution 
function of $\hat I$, from which we can calculate 
the values of the exponents \cite{Iwata2}. 
Since the calculation requires a complicated procedure, 
we present a method by which the values of the exponents
can be determined without the distribution function of $\hat I$.

The basic idea  of our analysis is to consider a distribution 
function of the average frequency $\hat \Omega$ over a time 
interval $\Dt=M I_*$, where $M $ is a large
number independent of $\epsilon$. (Note that $\Dt$ depends on 
$\epsilon.)$ For the explicit expression
\begin{equation}
\hat \Omega=\frac{1}{\Dt}\int_0^{\Dt} 
dt (\partial_t \theta),
\end{equation}
we can expect a large deviation property, which is given as
\begin{equation}
P(\hat \Omega=\Omega)\simeq \e^{- M  N G(\Omega)/T},
\label{rate:G}
\end{equation}
where the rate function $G(\Omega)$  takes a minimum value zero 
when $\Omega=\Omega_*$. 
Then, it can be shown that $I_*$ in (\ref{bi}) is equal to $2\pi/\Omega_*$.

We now estimate the rate function $G(\Omega)$.
Let $[\theta]$ be a trajectory $(\theta(t))_{t=0}^{\Delta t}$, and 
$\theta(0)$ is fixed as an arbitrary value. 
The probability density of trajectory is then expressed by 
\begin{equation}
{\cal P}([\theta])= \frac{1}{Z}\e^{
- \frac{N}{4T} \int_{0}^{\Dt}dt 
\left[(\p \theta-f(\theta))^2+\frac{2T}{N}f'(\theta) \right] },
\label{adis}
\end{equation}
where $f(\theta)=\omega-h\sin\theta$, the prime represents the
derivative with respect to $\theta$, and $Z$ is a normalization
factor. The last term corresponds to a Jacobian term associated with the 
transformation from a noise sequence 
$(\eta (t))_{t=0}^{\Delta t}$ to the trajectory $[\theta]$. By formally 
expressing $P(\hat \Omega=\Omega)$ as
\begin{equation}
P(\hat \Omega=\Omega)= \int {\cal D} [\theta]
{\cal P}([\theta]) \delta\left(\Omega-\frac{1}{\Dt}\int_0^{\Dt} 
dt (\partial_t \theta) \right),
\label{formal}
\end{equation} 
we consider the trajectory whose weight becomes most dominant in 
the limit $N \to \infty$. The trajectory, which is denoted
by $\thetas_\Omega$, is a periodic solution with period 
$2\pi/\Omega$ of the variational equation 
$\partial_t^2\theta(t) = -\partial_\theta U(\theta)/2$, where 
$U(\theta)=-f(\theta)^2$. The solution  $\thetas_\Omega(t)$ 
is obtained from the energy conservation equation, 
which leads to the derivation of 
\begin{equation}
\partial_t \thetas_\Omega= \sqrt{E(\Omega)-U(\thetas_\Omega)},
\label{thetas}
\end{equation}
where the parameter $E(\Omega)$ is related to the frequency $\Omega$ as 
\begin{equation}
\frac{2 \pi}{\Omega}= 
\int_0^{2 \pi}  
\frac{d \theta}{\sqrt{E(\Omega)-U(\theta)}}.
\label{Edef}
\end{equation}

Since $\thetas_\Omega$ contributes to $P(\hat \Omega=\Omega)$ 
much more than other $2\pi/\Omega$-periodic trajectories,  
it is reasonable to 
expect that $P(\hat{\Omega}=\Omega)\simeq {\cal P}([\thetas_\Omega])$. 
The substitution of (\ref{thetas}) into (\ref{adis}) yields
\begin{equation}
G(\Omega)=\frac{I_* \Omega}{8\pi} 
\int_0^{2 \pi} d\theta 
\frac{(\sqrt{E(\Omega)-U(\theta)}-\sqrt{-U(\theta)})^2}
{\sqrt{E(\Omega)-U(\theta)}}.
\end{equation}
It can be observed that $dG(\Omega)/d\Omega|_{\Omega_*}=0 $
and $G(\Omega_*)=0 $ when $\Omega_*$ satisfies the condition $E(\Omega_*)=0$. 
Therefore, the rate function
$G(\Omega)$ takes a quadratic form 
\begin{equation}
G(\Omega) = B(\epsilon) \Omega_*^{-4} (\Omega-\Omega_*)^2 
\label{G:approx}
\end{equation}
when $\Omega$ is close to $\Omega_*$, where 
$B(\epsilon)$ is calculated as 
\begin{equation}
B(\epsilon)=\frac{8\sqrt{2}\pi}{3}\epsilon^{5/2}+O(\epsilon^{7/2}).\label{B}
\end{equation}
Furthermore, by considering  (\ref{Edef}) with $E=0$, we obtain  
\begin{equation}
\Omega_* = \sqrt{2-\epsilon}\epsilon^{1/2}.
\label{omegas}
\end{equation}

Now, we consider the average of $\hat I$ during the time interval $\Dt$, 
which is denoted by $\hat J$. It can be easily confirmed that 
$\hat J = 2\pi/\hat \Omega$. 
Then, by the transformation of the variable in (\ref{rate:G}) and 
(\ref{G:approx}), we derive
\begin{equation}
P(\hat J=J)\simeq \e^{- M N  B(\epsilon)  
\left( J- 2 \pi/\Omega_* \right)^2/ (4 \pi^2 T)}.
\label{pj}
\end{equation}
By substituting (\ref{B}) and (\ref{omegas}) into (\ref{pj}),
we find that $\bra J \ket \simeq \epsilon^{-1/2}$ and 
$\bra (J- \bra J \ket) ^2 \ket \simeq \epsilon^{-5/2}$.
Since these $\epsilon$ dependences should be equal to those 
of $I_*$ and $\chi$, we arrive at the 
theoretical results $\zeta=1/2$ and $\gamma=5/2$.
These values coincide perfectly with the numerical values.

Furthermore, our analysis yields a new formula
for the phase diffusion constant $D$, 
which is expressed by $D\equiv\Dt \bra (\hat \Omega-\Omega_*)^2 \ket /2$
because $\hat \Omega=(\theta(\Dt)-\theta(0))/\Dt$. 
Indeed, from (\ref{rate:G}), we obtain   
\begin{equation}
D = \frac{T}{2N}  \frac{I_*}{G''(\Omega_*)},
\end{equation} 
which leads to the power-law  behavior 
$D =  (3T/32\sqrt{2}\pi N)\epsilon^{-1}+ O(\epsilon^0)$.
Here, with the crossover relation 
$\epsilon \simeq (N/T)^{-2/3}$,  we conjecture 
$D \simeq  (T/N)(N/T)^{2/3}$ at $\epsilon=0$,
which  was reported in Ref. \cite{Reimann}.

\paragraph{Concluding remarks:}        %

We have studied a simple model that exhibits 
critical behavior near a saddle-node bifurcation.
The power-law divergence, $\chi\simeq\epsilon^{-5/2}$, which we have predicted 
for coupled excitable elements will be observed in experimental systems.
Complicated systems such as those with a tactical network or 
integrate-and-fire dynamics will be analyzed by extending our theory.

The analysis of finite-dimensional systems is the next theoretical
problem.  
As usual in critical 
phenomena, we wish to determine the upper-critical dimension above which 
the values of the exponents are the same as those in the 
globally coupled model.
Then, we intend to develop a systematic method 
to take into account non-Gaussian fluctuations. 
The construction of such a theory is extremely interesting.

Before ending this Letter, 
let us recall that the amplitude of oscillation exhibits
a discontinuous transition at the saddle-node bifurcation.
Here, it should be noted that the co-existence of critical 
fluctuations with a discontinuous 
transition is one of the remarkable features of jamming
transitions \cite{Fisher}. 
This is not an accidental coincidence and can be 
explained in the following manner.

A standard characterization of the critical nature 
near a jamming transition is 
based on the nonlinear susceptibility $\chi_4(t)$, which 
quantifies the fluctuations of unlocking events during 
a time interval $t$ \cite{Onuki}. Among the several theories 
for $\chi_4(t)$ \cite{Biroli,Garrahan,Miyazaki,Iwata1}, 
one theory states that the divergence 
of $\chi_4(t)$ originates from the critical fluctuations of 
the time when an unlocking event occurs \cite{Iwata1}.  
By employing the method in Ref. \cite{Iwata1}, we can discuss
the divergent behavior of amplitude fluctuations in the present 
problem. That is,  the coexistence of a discontinuous transition 
and a critical fluctuation in coupled  excitable elements can be
described in a manner similar to  that in jamming transitions. 

Moreover, it has been recently shown that dynamical behaviors of the  
$k$-core percolation in a random graph exhibit a saddle-node bifurcation 
at the percolation point \cite{Iwata3}. 
Since it has been known that the $k$-core percolation is  
related to a kinetically constrained model and a random-field Ising model 
\cite{Duxbury,Schwarz,Toninelli,Dhar,Ohta}, our work might be useful 
for theoretical analysis of such systems. 

We hope that our theory of the nontrivial behavior of a simple model 
will stimulate further studies on subjects that increase the 
understanding of the cooperative nature of nonequilibrium systems. 

The authors thank M. Iwata for discussions on related problems. 
This work was supported by a grant from 
the Ministry of Education, Science, Sports and Culture of Japan, 
No. 19540394.

\end{document}